\documentclass[12pt,english]{article}
\usepackage{tgtermes}
\usepackage[T1]{fontenc}
\usepackage[latin9]{inputenc}
\usepackage[a4paper]{geometry}
\usepackage{array}
\usepackage{multirow}
\usepackage{amsmath}
\usepackage{graphicx}
\usepackage[authoryear]{natbib}

\makeatletter

\providecommand{\tabularnewline}{\\}

\makeatother

\usepackage{babel}
\begin{document}

\title{Adaptive MCMC for Generalized Method of\linebreak{}
Moments with Many Moment Conditions}

\author{Masahiro Tanaka\thanks{Graduate School of Economics, Waseda University; Director-General
for Policy Planning (Statistical Standards), Ministry of Internal
Affairs and Communications. Address: 1-6-1, Nishi-Waseda, Shinjuku-ku,
Tokyo 169-8050 Japan, Email: gspddlnit45@toki.waseda.jp.}}

\date{March 10, 2021}
\maketitle
\begin{abstract}
A generalized method of moments (GMM) estimator is unreliable for
a large number of moment conditions, that is, it is comparable, or
larger than the sample size. While classical GMM literature proposes
several provisions to this problem, its Bayesian counterpart (i.e.,
Bayesian inference using a GMM criterion as a quasi-likelihood) almost
totally ignores it. This study bridges this gap by proposing an adaptive
Markov Chain Monte Carlo (MCMC) approach to a GMM inference with many
moment conditions. Particularly, this study focuses on the adaptive
tuning of a weighting matrix on the fly. Our proposal consists of
two elements. The first is the use of the nonparametric eigenvalue-regularized
precision matrix estimator, which contributes to numerical stability.
The second is the random update of a weighting matrix, which substantially
reduces computational cost, while maintaining the accuracy of the
estimation. We then present a simulation study and real data application
to compare the performance of the proposed approach with existing
approaches.

\bigskip{}


\end{abstract}

\section{Introduction}

The generalized method of moments (GMM) is a widely used statistical
framework \citep{Hansen1982,Hall2005}. It estimates unknown parameters
via a set of moment conditions. A parameter estimate is obtained by
minimizing a GMM criterion constructed as a quadratic form and composed
of the sample mean of a vector-valued function that represents the
moment conditions and a weighting matrix. While GMM uses only lower-order
moments, thus, being statistically less efficient than full-information
methods such as the maximum likelihood method, it has many advantages,
including robustness to model misspecification, nonparametric treatment
of heteroskedasticity, and computational simplicity. 

This study focuses on the Bayesian version of GMM. A GMM criterion
can be viewed as a quasi-likelihood, being theoretically equivalent
to the Laplace approximation of the true likelihood around its mode
\citep{Chernozhukov2003}. Exploiting this feature, one can conduct
a (quasi-)Bayesian inference by replacing true likelihood with a GMM
criterion, as discussed by, for example, \citet{Kim2002,Yin2009}.\footnote{See also \citet{Belloni2009,Li2016} for a discussion of theoretical
properties. } Posterior draws from a quasi-posterior density (product of quasi-likelihood
and prior density) can be simulated using standard Bayesian Markov
Chain Monte Carlo (MCMC) techniques, such as the Metropolis-Hastings
algorithm. In this study, we call this inferential approach Bayesian
GMM, in contradistinction to classical GMM.

A GMM criterion has many moment conditions for applications, making
the estimator considerably unreliable. In some cases, the number of
moment conditions can be large, including dynamic panel models (e.g.,
\citealp{Arellano1991,Blundell1998,Roberts2009,Vieira2012}), instrumental
variable methods (e.g., \citealp{Chernozhukov2005,Chernozhukov2013}),
and identification through heteroskedasticity \citep{Lewbel2012}. 

The literature on classical GMM proposes several provisions to the
problem, such as systematic moment selection \citep{Andrews1999,Andrews2001,Hall2003,Hall2007,Okui2009,Donald2009,Canay2010,DiTraglia2016,Chang2018},
averaging \citep{Chen2016}, and shrinkage estimation \citep{Liao2013,Fan2014,Cheng2015,Caner2018}.
On the contrary, the literature on Bayesian GMM largely ignores the
problem, although remedies tailored to classical GMM are not straightforwardly
applicable to Bayesian GMM for two reasons. First, they are two-stage
procedures in which the final estimate is computed based on the first
estimate with the identity weighting matrix. However, such a strategy
is not feasible in Bayesian GMM, because the relative contributions
of a GMM criterion (quasi-likelihood) and a prior density to the quasi-posterior
depend on the weighting matrix, the mode of a quasi-posterior under
the identity weighting matrix is not consistent with that under the
optimal weighting matrix. Therefore, in Bayesian GMM, a weighting
matrix has to be estimated with the unknown parameters of interest.
Second, Bayesian GMM is often used in cases where numerical optimization
does not work well because a GMM criterion is discontinuous in parameters
or has many local optima. Therefore, even when a non-informative prior
is employed, in some cases, a first-step estimate is not readily available.
The purpose of this study is to bridge this gap by proposing a novel
method to deal with Bayesian GMM with many moment conditions. 

For both classical and Bayesian GMM, choosing a good weighting matrix
is a significant issue. It is theoretically optimal to set a weighting
matrix to the precision matrix (i.e., the inverse of the covariance
matrix) of moment conditions, evaluated based on true parameter values.
As this approach is infeasible in practice, two-step and continuously
updated estimators are commonly used in classical GMM \citep{Hansen1982,Hansen1996}.
In contrast, the literature on Bayesian GMM focuses less on the weighting
matrix choice. \citet{Chernozhukov2003}, who use the random-walk
Metropolis-Hasting algorithm, suggest recomputing the weighting matrix
each time a parameter proposal is drawn; a posterior mean of the weighting
matrix is supposed to be optimal on average. In this approach, the
unknown parameters and a weighting matrix are updated concurrently.
Consequently, the surface of the quasi-posterior becomes complicated,
making the MCMC algorithm inefficient and unstable. To tackle this
problem, \citet{Yin2011} propose an approach they call the stochastic
GMM, in which unknown parameters are updated consecutively, and the
corresponding weighting matrix is also updated accordingly. Their
approach improves the numerical stability of the posterior simulator
by suppressing changes in the posterior in a single cycle. However,
this approach requires so many matrix inversions of the weighting
matrix that it is not practical for models with many moment conditions.

There are two difficulties in setting a weighting matrix for a large
number of moment conditions. First, as in classical GMM, the sample
estimate of the covariance matrix of the moment conditions is unreliable,
and the inversion of the covariance matrix can amplify estimation
errors. Second, it is computationally demanding because the inversion
of the sample covariance matrix is repeatedly computed. This problem
is specific to Bayesian GMM. 

In this study, we develop an adaptive MCMC approach to tackle the
problem of many moment conditions in Bayesian GMM. The proposal consists
of two main contributions. First, we propose estimating the precision
matrix of the moment conditions using the nonparametric eigenvalue-regularized
precision matrix estimator developed by \citet{Lam2016}. This estimator
is more numerically stable than the standard estimator, the inverse
of a sample covariance matrix. Through a series of Monte Carlo experiments,
we show that the proposed approach outperforms existing ones in terms
of both statistical and computational efficiency. Second, we propose
a random updating of a weighting matrix using the recursive mean of
the posterior samples. In our approach, we set adaptation probabilities
to decrease exponentially, which ensures the validity of the MCMC
algorithm, and significantly saves computational cost.

This paper proceeds as follows: Section 2 introduces the proposed
approach. Section 3 conducts a simulation study. In Section 4, we
apply the approach to a real data problem as an example. Section 5
concludes this paper with a discussion.

\section{Method}

\subsection{Setup and challenges}

We consider the Bayesian inference of a statistical model using a
set of moment conditions. Let us assume that a likelihood function
can be approximated by a quasi-likelihood based on a GMM criterion
\citep{Hansen1982}. We call this inferential approach Bayesian GMM
\citep{Kim2002,Yin2009}. Given data $\mathcal{D}$ and an $L$-dimensional
parameter $\boldsymbol{\theta}$, we estimate a statistical model
through a set of moment conditions represented by a $K$-dimensional
vector of moment functions $\boldsymbol{m}_{n}\left(\boldsymbol{\theta}\right)=\left(m_{n,1}\left(\boldsymbol{\theta}\right),...,m_{n,K}\left(\boldsymbol{\theta}\right)\right)$:
\[
E\left[\boldsymbol{m}_{n}\left(\boldsymbol{\theta}\right)\right]=\boldsymbol{0}_{K}.
\]
A GMM criterion $v\left(\boldsymbol{\theta}\right)$ is defined as
the quadratic form of the sample mean of $\boldsymbol{m}_{n}\left(\boldsymbol{\theta}\right)$,
denoted by $\bar{\boldsymbol{m}}\left(\boldsymbol{\theta}\right)$,
and a symmetric positive definite weighting matrix $\boldsymbol{W}$:
\[
v\left(\boldsymbol{\theta}\right)=\bar{\boldsymbol{m}}\left(\boldsymbol{\theta}\right)^{\top}\boldsymbol{W}\bar{\boldsymbol{m}}\left(\boldsymbol{\theta}\right),
\]
\[
\bar{\boldsymbol{m}}\left(\boldsymbol{\theta}\right)=\frac{1}{N}\sum_{n=1}^{N}\boldsymbol{m}_{n}\left(\boldsymbol{\theta}\right),
\]
where $N$ is the sample size. For notational convenience, we omit
the dependence on $\mathcal{D}$ from functions $\boldsymbol{m}_{n}\left(\boldsymbol{\theta}\right)$,
$\bar{\boldsymbol{m}}\left(\boldsymbol{\theta}\right)$, and $v\left(\boldsymbol{\theta}\right)$.
A quasi-likelihood is defined based on the GMM criterion as 

\begin{eqnarray*}
q\left(\boldsymbol{\theta}|\mathcal{D}\right) & = & \left(\frac{2\pi}{N}\right)^{-\frac{K}{2}}\det\left(\boldsymbol{W}\right)^{\frac{1}{2}}\exp\left[-\frac{N}{2}v\left(\boldsymbol{\theta}\right)\right]\\
 & = & \left(\frac{2\pi}{N}\right)^{-\frac{K}{2}}\det\left(\boldsymbol{W}\right)^{\frac{1}{2}}\exp\left[-\frac{N}{2}\bar{\boldsymbol{m}}\left(\boldsymbol{\theta}\right)^{\top}\boldsymbol{W}\bar{\boldsymbol{m}}\left(\boldsymbol{\theta}\right)\right].
\end{eqnarray*}
A GMM criterion can be seen as the Laplace approximation of the negative
true likelihood evaluated around the mode \citep{Chernozhukov2003}.
Given a prior density $p\left(\boldsymbol{\theta}\right)$, the posterior
density $p\left(\boldsymbol{\theta}|\mathcal{D}\right)$ is approximated
as

\begin{equation}
p\left(\boldsymbol{\theta}|\mathcal{D}\right)\approx\frac{q\left(\boldsymbol{\theta}|\mathcal{D}\right)p\left(\boldsymbol{\theta}\right)}{\int q\left(\boldsymbol{\theta}^{\prime}|\mathcal{D}\right)p\left(\boldsymbol{\theta}^{\prime}\right)d\boldsymbol{\theta}^{\prime}},\label{eq: posterior}
\end{equation}
where the denominator is generally unknown but constant. The posterior
samples $\boldsymbol{\theta}_{\left[j\right]}=$$\left(\theta_{\left[j\right],1},...,\theta_{\left[j\right],L}\right)^{\top}$
are drawn from this target density (evaluated up to the normalizing
constant) using Bayesian simulation techniques. We use the Metropolis-Hastings
(MH) algorithm similar to previous studies (e.g., \citealp{Chernozhukov2003,Yin2009}).
Given a current state $\boldsymbol{\theta}$, a single step of a MH
algorithm is specified as follows:
\begin{enumerate}
\item A proposal $\boldsymbol{\theta}^{\prime}$ is generated from a proposal
kernel $p\left(\boldsymbol{\theta}^{\prime}|\boldsymbol{\theta}\right)$. 
\item Compute the MH ratio $\alpha\left(\boldsymbol{\theta}^{\prime},\boldsymbol{\theta}\right)$
as 
\[
\alpha\left(\boldsymbol{\theta}^{\prime},\boldsymbol{\theta}\right)=\frac{q\left(\boldsymbol{\theta}^{\prime}|\mathcal{D}\right)p\left(\boldsymbol{\theta}^{\prime}\right)p\left(\boldsymbol{\theta}^{\prime}|\boldsymbol{\theta}\right)}{q\left(\boldsymbol{\theta}|\mathcal{D}\right)p\left(\boldsymbol{\theta}\right)p\left(\boldsymbol{\theta}|\boldsymbol{\theta}^{\prime}\right)}.
\]
\item Set the next state to the proposal $\boldsymbol{\theta}^{*}=\boldsymbol{\theta}^{\prime}$
with probability of $\alpha\left(\boldsymbol{\theta}^{\prime},\boldsymbol{\theta}\right)\lor1$,
or set the next state to the current state $\boldsymbol{\theta}^{*}=\boldsymbol{\theta}$
with probability of $1-\left(\alpha\left(\boldsymbol{\theta}^{\prime},\boldsymbol{\theta}\right)\lor1\right)$.
\item Return the next state $\boldsymbol{\theta}^{*}$.
\end{enumerate}
As in classical GMM, the statistical efficiency of the Bayesian GMM
critically depends on the choice of the weighting matrix $\boldsymbol{W}$.
$\boldsymbol{W}$ is optimal when it is set to the precision matrix
of the moment conditions based on the true parameter values $\boldsymbol{\theta}_{0}$.
This choice is optimal in that it minimizes the Kullback-Leibler divergence
of the true data generating process to the set of all asymptotically
less restrictive distributions \citep{Li2016}. Let $\boldsymbol{M}\left(\boldsymbol{\theta}\right)=\left(\boldsymbol{m}_{1}\left(\boldsymbol{\theta}\right),...,\boldsymbol{m}_{N}\left(\boldsymbol{\theta}\right)\right)^{\top}$
denote an $N$-by-$K$ matrix of the moment functions. The optimal
choice of weighting matrix in finite sample is 
\[
\boldsymbol{W}\left(\boldsymbol{\theta}_{0}\right)=\left[N^{-1}\boldsymbol{M}\left(\boldsymbol{\theta}_{0}\right)^{\top}\boldsymbol{M}\left(\boldsymbol{\theta}_{0}\right)\right]^{-1}.
\]
It is a common practice in classical GMM to employ the two-step \citep{Hansen1982}
or continuously updating estimators \citep{Hansen1996}. The two-step
estimation method obtains a first-stage estimate using an arbitrary
weighting matrix (e.g., an identity matrix), then obtains a second-stage
estimate using a weighting matrix to a precision matrix of the moment
conditions based on the first-stage estimate. The continuously updating
estimation method repeats the two-step estimation for more than once. 

Despite its critical importance, few studies have examined the practical
choice of $\boldsymbol{W}$ in the context of Bayesian GMM. A straightforward
approach to choosing $\boldsymbol{W}$, which is employed, for instance,
by \citet{Chernozhukov2003,Yin2009}, can be described as follows:
At the $j$th MCMC iteration, given the current parameters $\boldsymbol{\theta}_{\left[j-1\right]}$,
a proposal $\boldsymbol{\theta}^{\prime}$ is simulated for a proposal
density $p\left(\boldsymbol{\theta}^{\prime}|\boldsymbol{\theta}_{\left[j-1\right]}\right)$.
For simplicity, we assume the density is symmetric, for example, a
normal distribution. The weighting matrix is set to the precision
matrix of the moment condition based on $\boldsymbol{\theta}^{\prime}$,
that is, the parameter vector and weighting matrix are concurrently
proposed and updated (i.e., accepted or rejected). We call this approach
the concurrent GMM. The MH ratio is calculated as
\begin{eqnarray*}
\alpha\left(\boldsymbol{\theta}^{\prime},\boldsymbol{\theta}_{\left[j-1\right]}\right) & = & \frac{q\left(\boldsymbol{\theta}^{\prime}|\mathcal{D}\right)p\left(\boldsymbol{\theta}^{\prime}\right)p\left(\boldsymbol{\theta}^{\prime}|\boldsymbol{\theta}_{\left[j-1\right]}\right)}{q\left(\boldsymbol{\theta}_{\left[j-1\right]}|\mathcal{D}\right)p\left(\boldsymbol{\theta}_{\left[j-1\right]}\right)p\left(\boldsymbol{\theta}_{\left[j-1\right]}|\boldsymbol{\theta}^{\prime}\right)}\\
 & = & \frac{\det\left(\boldsymbol{W}\left(\boldsymbol{\theta}^{\prime}\right)\right)^{\frac{1}{2}}\exp\left[-\frac{N}{2}\bar{\boldsymbol{m}}\left(\boldsymbol{\theta}^{\prime}\right)^{\top}\boldsymbol{W}\left(\boldsymbol{\theta}^{\prime}\right)\bar{\boldsymbol{m}}\left(\boldsymbol{\theta}^{\prime}\right)\right]p\left(\boldsymbol{\theta}^{\prime}\right)}{\det\left(\boldsymbol{W}\left(\boldsymbol{\theta}_{\left[j-1\right]}\right)\right)^{\frac{1}{2}}\exp\left[-\frac{N}{2}\bar{\boldsymbol{m}}\left(\boldsymbol{\theta}_{\left[j-1\right]}\right)^{\top}\boldsymbol{W}\left(\boldsymbol{\theta}_{\left[j-1\right]}\right)\bar{\boldsymbol{m}}\left(\boldsymbol{\theta}_{\left[j-1\right]}\right)\right]p\left(\boldsymbol{\theta}_{\left[j-1\right]}\right)}.
\end{eqnarray*}
This approach is motivated by setting a weighting matrix to an optimal
one on average. Note that uncertainty about $\boldsymbol{W}$ is inherently
different from that about $\boldsymbol{\theta}$; $\boldsymbol{W}$
is not inferred using a prior but it is crudely tuned along the posterior
simulation. 

\citet{Yin2011} argue this approach is slow to converge, because
the concurrent updating of $\boldsymbol{\theta}$ and $\boldsymbol{W}$
complicates the surface of the target density, resulting in an inefficient
move of the MH sampler. They propose an alternative approach, stochastic
GMM, where the elements of $\boldsymbol{\theta}$ are updated one
by one, keeping $\boldsymbol{W}$ unchanged. This approach is designed
to update $\boldsymbol{\theta}$ and $\boldsymbol{W}$ gradually,
suppressing instantaneous changes in the shape of the target density.
Let $\boldsymbol{\theta}_{\left[j,l\right]}=\left(\theta_{\left[j,l\right],1},...,\theta_{\left[j,l\right],l},\theta_{\left[j,l-1\right],l+1},...,\theta_{\left[j,l-1\right],L}\right)^{\top}$
denote a state at the $j$th MCMC iteration after the $l$th parameter
was updated. Once a proposed value of $\theta_{\left[j,l\right],l}^{\prime}$
is simulated, a proposal is constructed as $\boldsymbol{\theta}_{\left[j,l\right]}^{\prime}=\left(\theta_{\left[j,l\right],1},...,\theta_{\left[j,l\right],l-1},\theta_{\left[j,l\right],l}^{\prime},\theta_{\left[j.l-1\right],l+1},...,\theta_{\left[j,l-1\right],L}\right)^{\top}$,
and the MH ratio is given by 
\[
\alpha\left(\boldsymbol{\theta}_{\left[j,l\right]}^{\prime},\boldsymbol{\theta}_{\left[j,l-1\right]}\right)=\frac{\exp\left[-\frac{N}{2}\bar{\boldsymbol{m}}\left(\boldsymbol{\theta}_{\left[j,l\right]}^{\prime}\right)^{\top}\boldsymbol{W}\left(\boldsymbol{\theta}_{\left[j,l-1\right]}\right)\bar{\boldsymbol{m}}\left(\boldsymbol{\theta}_{\left[j.l\right]}^{\prime}\right)\right]p\left(\boldsymbol{\theta}_{\left[j.l\right]}^{\prime}\right)}{\exp\left[-\frac{N}{2}\bar{\boldsymbol{m}}\left(\boldsymbol{\theta}_{\left[j,l-1\right]}\right)^{\top}\boldsymbol{W}\left(\boldsymbol{\theta}_{\left[j,l-1\right]}\right)\bar{\boldsymbol{m}}\left(\boldsymbol{\theta}_{\left[j,l-1\right]}\right)\right]p\left(\boldsymbol{\theta}_{\left[j,l-1\right]}\right)}.
\]
The underlying justification of this approach is the same as the concurrent
GMM. This approach is computationally heavy for a large number of
moment conditions, because it requires many matrix inversions.

There are two challenges in the choice of the weighting matrix for
Bayesian GMM, especially when the number of moment conditions $K$
is large, that is, $K$ is comparable to or even larger than the sample
size $N$. First, when $K$ is large, the covariance of the moment
functions is ill-estimated, and estimation errors are amplified through
matrix inversions. As mentioned in Section 2.1, remedies in classical
GMM literature cannot be directly imported to Bayesian GMM. A simple
solution is using the Moore-Penrose generalized inverse, but it does
not work well, as the simulation study in Section 2.3 shows.\footnote{See \citet{Satchachai2008} on this point for classical GMM.}\footnote{In classical GMM, \citet{Doran2006} suggest using the principal components
of a weighting matrix. From the author's experience, a strategy using
the standard principal component analysis to estimate the weighting
matrix does not work well for Bayesian GMM, which this study does
not consider.} The second challenge is the computational cost. The existing approaches
require repeated inversion of the sample covariance of the moment
functions, thus, imposing severe computational loads. 

\subsection{Proposed approach}

Our proposal comprises two elements: regularized precision matrix
estimation and random update of the weighting matrix. The former aims
to improve the numerical stability in the update of $\boldsymbol{W}$,
while the latter is introduced to reduce the computational cost.

First, we propose to compute $\boldsymbol{W}$ using the nonparametric
eigenvalue-regularized (NER) precision matrix estimator \citep{Lam2016},
in which the eigenvalues of a sample covariance matrix are regularized
through the splitting of data.\footnote{\citet{Abadir2014} consider a closely related covariance estimator.}
The estimator has several favorable properties. First, it is asymptotically
optimal with respect to Stein's loss (Proposition 2 in \citealp{Lam2016},
p. 937). Second, it is optimization-free, and thus, computationally
less demanding than the other shrinkage covariance/precision matrix
estimators.\footnote{See, \citet{Pourahmadi2011,Fan2016,Lam2020} for a survey of the literature
on covariance/precision matrix estimation.}

Given $\boldsymbol{\theta}$, the moment functions are partitioned
as $\boldsymbol{M}\left(\boldsymbol{\theta}\right)=\left(\boldsymbol{M}_{1}\left(\boldsymbol{\theta}\right)^{\top},\;\boldsymbol{M}_{2}\left(\boldsymbol{\theta}\right)^{\top}\right)^{\top}$,
where the sizes of $\boldsymbol{M}_{1}\left(\boldsymbol{\theta}\right)$
and $\boldsymbol{M}_{2}\left(\boldsymbol{\theta}\right)$ are $N_{1}$-by-$K$
and $N_{2}$-by-$K$, respectively. The covariance matrices of the
sub-samples are computed in a standard manner: $\tilde{\boldsymbol{\Sigma}}_{i}=N_{i}^{-1}\boldsymbol{M}_{i}\left(\boldsymbol{\theta}\right)^{\top}\boldsymbol{M}_{i}\left(\boldsymbol{\theta}\right)$,
$i=1,2$. Let $N^{*}$ denote the sample size of the first sub-sample,
or the splitting location, $N_{1}=N^{*}$, and then, $N_{2}=N-N^{*}$.
The eigenvalue decomposition of $\tilde{\boldsymbol{\Sigma}}_{i}$
is represented by $\tilde{\boldsymbol{\Sigma}}_{i}=\boldsymbol{P}_{i}\boldsymbol{D}_{i}\boldsymbol{P}_{i}^{\top}$,
$i=1,2$, where $\boldsymbol{D}_{i}=\textrm{diag}\left(d_{i,1},...,d_{i,K}\right)$
is a diagonal matrix containing the eigenvalues of $\tilde{\boldsymbol{\Sigma}}_{i}$,
$d_{i,1}\geq\cdots\geq d_{i,K}$, and $\boldsymbol{P}_{i}=\left(\boldsymbol{p}_{i,1},...,\boldsymbol{p}_{i,K}\right)$
is a matrix composed of the corresponding eigenvectors. Following
\citet{Lam2016}, the sample covariance matrix of the moment functions
is estimated as
\[
\tilde{\boldsymbol{\Sigma}}_{NER}=\boldsymbol{P}_{1}\left[\left(\boldsymbol{P}_{1}^{\top}\tilde{\boldsymbol{\Sigma}}_{2}\boldsymbol{P}_{1}\right)\odot\boldsymbol{I}_{K}\right]\boldsymbol{P}_{1}^{\top},
\]
where $\boldsymbol{I}_{K}$ is a $K$-dimensional identity matrix
and $\odot$ denotes the Hadamard product. Therefore, the corresponding
precision matrix is given by 
\begin{equation}
\boldsymbol{W}\left(\boldsymbol{\theta}\right)=\boldsymbol{P}_{1}\left[\left(\boldsymbol{P}_{1}^{\top}\tilde{\boldsymbol{\Sigma}}_{2}\boldsymbol{P}_{1}\right)\odot\boldsymbol{I}_{K}\right]^{-1}\boldsymbol{P}_{1}^{\top}.\label{eq: NER}
\end{equation}
\citet{Lam2016} suggests improving this estimator by averaging many
(e.g., 50) estimates, using different sets of partitioned data that
are generated via random permutation. For robustness, we also randomly
permute $\boldsymbol{m}_{n}\left(\boldsymbol{\theta}\right)$, $n=1,...,N$,
for each computation of $\boldsymbol{W}$. 

The choice of the split location $N^{*}$ is non-trivial. Theorem
5 of \citeauthor{Lam2016} (2016, p. 941) suggests that when $K/N\rightarrow c$,
it is asymptotically efficient to choose $N^{*}=N-aN^{1/2}$, with
some constants $c,a>0$. However, this poses two difficulties. First,
this asymptotic property is not applicable when $N^{*}/N$ goes to
a constant smaller than 1. Second, there is no practical guidance
for setting $a$. \citet{Lam2016} proposes to choose $N^{*}$ to
minimize the following criterion using a grid search:
\begin{equation}
g\left(N^{*}\right)=\left\Vert \sum_{m=1}^{M}\left(\tilde{\boldsymbol{\Sigma}}_{NER}^{\left(m\right)}-\tilde{\boldsymbol{\Sigma}}_{2}^{\left(m\right)}\right)\right\Vert _{F}^{2},\label{eq: fro-norm crit}
\end{equation}
where the superscripts for $\boldsymbol{\Sigma}_{NER}^{\left(m\right)}$
and $\boldsymbol{\Sigma}_{2}^{\left(m\right)}$ denote indices for
different permutations, $M$ is the number of permutations executed,
and $\left\Vert \cdot\right\Vert _{F}$ denotes the Frobenius norm.
\citet{Lam2016} considers the following grid as a set of candidates
for $N^{*}$:
\begin{equation}
\left\{ 2N^{1/2},\;0.2N,\;0.4N,\;0.6N,\;0.8N,\;N-2.5N^{1/2},\;N-1.5N^{1/2}\right\} .\label{eq: grid bigN_star}
\end{equation}
In our framework, one might consider tuning $N^{*}$ adaptively based
on the above criterion. However, we do not adopt such a strategy,
because the criterion is not informative enough to pin down the optimal
choice of $N^{*}$, as the subsequent section shows. A default choice
in this study is $N^{*}=0.6N$, that is, the median of \citeauthor{Lam2016}'s
(2016) grid. As the next section shows, simulated posteriors are not
sensitive to $N^{*}$, as long as $N^{*}$ is within a moderate range.

Next, we consider randomly updating a weighting matrix $\boldsymbol{W}$.
We explicitly treat $\boldsymbol{W}$ as a tuning parameter, and update
it on the fly, as in adaptive MCMC algorithms \citep{Haario2001,Andrieu2008,Roberts2009}.
Our adaptation procedure is motivated by \citet{Bhattacharya2011}.
At the $j$th MCMC iteration, the adaptation of $\boldsymbol{W}$
occurs with probability $s\left(j\right)=\exp\left(\alpha_{0}+\alpha_{1}j\right)$,
regardless of the previous proposal being accepted or rejected. Throughout
the study, we chose $\alpha_{0}=-1$ and $\alpha_{1}=-10/J_{warmup}$,
where $J_{warmup}$ denotes the number of warmup iterations. If an
adaptation occurs, $\boldsymbol{W}$ is updated using the mean of
the previous sample obtained; at the $j$th iteration, $\bar{\boldsymbol{\theta}}_{\left[j-1\right]}=\left(j-1\right)^{-1}\sum_{j^{\prime}=1}^{j-1}\boldsymbol{\theta}_{\left[j^{\prime}\right]}$.
After warmup iterations, $\boldsymbol{W}$ is fixed to the end. This
adaptation strategy satisfies the convergence condition in Theorem
5 of \citet{Roberts2007}. In our implementation, at every $j$th
iteration, a random variable is simulated from a standard uniform
distribution, $u_{j}\sim\mathcal{U}\left(0,1\right)$, and $\boldsymbol{W}$
is updated if $u_{j}<s\left(j\right)$, where $\mathcal{U}\left(a,b\right)$
denotes a uniform distribution with support on interval $\left(a,b\right)$.
At the $j$th iteration, given a proposal $\boldsymbol{\theta}^{\prime}$,
the MH ratio is calculated as
\[
\alpha\left(\boldsymbol{\theta}^{\prime},\boldsymbol{\theta}_{\left[j-1\right]}\right)=\frac{\exp\left[-\frac{N}{2}\bar{\boldsymbol{m}}\left(\boldsymbol{\theta}^{\prime}\right)^{\top}\boldsymbol{W}\left(\bar{\boldsymbol{\theta}}_{\left[j-1\right]}\right)\bar{\boldsymbol{m}}\left(\boldsymbol{\theta}^{\prime}\right)\right]p\left(\boldsymbol{\theta}^{\prime}\right)}{\exp\left[-\frac{N}{2}\bar{\boldsymbol{m}}\left(\boldsymbol{\theta}_{\left[j-1\right]}\right)^{\top}\boldsymbol{W}\left(\bar{\boldsymbol{\theta}}_{\left[j-1\right]}\right)\bar{\boldsymbol{m}}\left(\boldsymbol{\theta}_{\left[j-1\right]}\right)\right]p\left(\boldsymbol{\theta}_{\left[j-1\right]}\right)}.
\]
This treatment of $\boldsymbol{W}$ does not conflict with the theoretical
results of Bayesian GMM (e.g., \citealt{Kim2002,Chernozhukov2003,Belloni2009,Li2016}).
Although the existing literature contains a discrepancy between theory
and practical computation in the treatment of a weighting matrix,
our treatment of $\boldsymbol{W}$ agrees better with the theoretical
results than do existing approaches. 

\section{Simulation Study}

We compare the proposed approach with alternatives.\footnote{The programs in this study are written in Matlab 2019b (64bit), and
executed on an Ubuntu Desktop 18.04 LTS (64bit), running on AMD Ryzen
Threadripper 1950X (4.2GHz).} We compare the NER estimator given by (\ref{eq: NER}) with the standard
estimators specified by
\[
\boldsymbol{W}\left(\boldsymbol{\theta}\right)=\begin{cases}
\left[N^{-1}\boldsymbol{M}\left(\boldsymbol{\theta}\right)^{\top}\boldsymbol{M}\left(\boldsymbol{\theta}\right)\right]^{-1}, & K\leq N,\\
\left[N^{-1}\boldsymbol{M}\left(\boldsymbol{\theta}\right)^{\top}\boldsymbol{M}\left(\boldsymbol{\theta}\right)\right]^{+}, & K>N,
\end{cases}
\]
where $\boldsymbol{A}^{+}$ denotes the Moore-Penrose generalized
inverse of a matrix $\boldsymbol{A}$. We consider six adaptation
strategies. The first is fixing the weighting matrix of the moment
conditions based on the true parameter value (\textit{Oracle}), the
second is the concurrent Bayesian GMM (\textit{Concurrent}) \citep{Chernozhukov2003,Yin2009},
and the third is the stochastic GMM (\textit{Stochastic}) \citep{Yin2011}.
The fourth is an MCMC version of the continuously updating GMM estimator
\citep{Hansen1996} (\textit{Continuous}), that is, $\boldsymbol{W}$
is updated in each cycle based on the current recursive means of the
sampled parameters. The fifth is the random update strategy we propose
(\textit{Random}).

We adopt an instrumental variable (IV) regression as the laboratory.
A true data generating process is specified by the following equations,
for $n=1,...,N$, 

\begin{equation}
x_{n}=\boldsymbol{z}_{n}^{\top}\boldsymbol{\delta}+w_{n},\quad w_{n}\sim\mathcal{N}\left(0,\sigma_{x}^{2}\right),\label{eq: IV1}
\end{equation}
\begin{equation}
y_{n}=\gamma x_{n}+\varphi\left(x_{n}-\boldsymbol{z}_{n}^{\top}\boldsymbol{\delta}\right)+u_{n},\quad u_{n}\sim\mathcal{N}\left(0,\sigma_{y}^{2}\right),\label{eq: IV2}
\end{equation}
where $y_{n}$ is a response variable, $x_{n}$ is an endogenous covariate,
$\boldsymbol{z}_{n}$ is a $K$-dimensional vector of instruments,
$u_{n}$ and $w_{n}$ are normally distributed errors, and $\mathcal{N}\left(\mu,\sigma^{2}\right)$
denotes a normal distribution with mean $\mu$ and variance $\sigma^{2}$.
$\gamma=0.5$ is a coefficient to be inferred. $\varphi=0.2$ is a
fixed parameter. The instruments are generated from a latent factor
model: for $n=1,...,N$,
\[
\boldsymbol{z}_{n}=\boldsymbol{B}\boldsymbol{\nu}_{n}+\boldsymbol{\epsilon}_{n},
\]
\[
\boldsymbol{\nu}_{n}\sim\mathcal{N}\left(\boldsymbol{0}_{S},\boldsymbol{I}_{S}\right),\quad\boldsymbol{\epsilon}_{n}\sim\mathcal{N}\left(\boldsymbol{0}_{K},\boldsymbol{\Psi}^{2}\right),
\]
where $S$ is the number of latent factors, $\boldsymbol{\nu}_{n}$
is an $S$-dimensional vector of latent factors, $\boldsymbol{\epsilon}_{n}$
is a $K$-dimensional vector of idiosyncratic errors with covariance
$\boldsymbol{\Psi}^{2}$, and $\boldsymbol{B}$ is a $K$-by-$S$
matrix of factor loadings. The distribution of $\boldsymbol{z}_{n}$
is written as 
\[
\boldsymbol{z}_{n}\sim\mathcal{N}\left(\boldsymbol{0}_{K},\;\boldsymbol{B}\boldsymbol{B}^{\top}+\boldsymbol{\Psi}^{2}\right),\quad n=1,...,N.
\]
 $\boldsymbol{\Psi}^{2}$ and $\boldsymbol{B}$ are set as follows:

\[
\boldsymbol{\Psi}^{2}=\textrm{diag}\left(\psi_{1}^{2},...,\psi_{K}^{2}\right),\quad\psi_{k}\sim\mathcal{U}\left(2,4\right),\quad k=1,...,K,
\]
\[
\boldsymbol{B}=\left(b_{k,s}\right),\quad b_{k,s}\sim\mathcal{U}\left(0,1\right),\quad k=1,...,K;\;s=1,...,S.
\]
The coefficients of $\boldsymbol{z}_{n}$ are generated as 

\[
\boldsymbol{\delta}=\boldsymbol{A}^{\top}\boldsymbol{\eta},\quad\boldsymbol{A}=\boldsymbol{B}^{\top}\left(\boldsymbol{B}\boldsymbol{B}^{\top}+\boldsymbol{\Psi}^{2}\right)^{-1},
\]
\[
\boldsymbol{\eta}=\left(\eta_{1},...,\eta_{S}\right)^{\top},\quad\eta_{s}\sim\mathcal{U}\left(0,1\right),\quad s=1,...,S.
\]
We consider three scenarios with different numbers of instruments
$K=\left\{ 50,150,250\right\} $ and factors $S=\left\{ K,K/2,3\right\} $.
We choose the standard deviations of the errors, $\sigma_{x}$ and
$\sigma_{y}$, so that the ratios of the standard deviations of the
errors to those of the signals, denoted by $q_{x}$ and $q_{y}$,
are $\varsigma_{x}$ and $\varsigma_{y}$, respectively: 
\[
\sigma_{x}=\varsigma_{x}q_{x},\quad\sigma_{y}=\varsigma_{y}q_{y},
\]
\[
q_{x}=\sqrt{\boldsymbol{\delta}^{\top}\left(\boldsymbol{B}\boldsymbol{B}^{\top}+\boldsymbol{\Psi}^{2}\right)\boldsymbol{\delta}},
\]
\[
q_{y}=\sqrt{\gamma^{2}\left(1+\varsigma_{x}^{2}\right)+\varphi^{2}\varsigma_{x}^{2}}q_{x}.
\]
We fix $\varsigma_{x}=\varsigma_{y}=2$. We infer the unknown parameter
$\gamma$ through a set of moment conditions, 
\[
E\left[\left(y_{n}-\gamma x_{n}\right)\boldsymbol{z}_{n}\right]=\boldsymbol{0}_{K}.
\]
We assign a flat prior on $\gamma$, $p\left(\gamma\right)\propto1$.
The sample size is fixed at $N=200$. For posterior sampling, we employ
an adaptive MH sampler of \citet{Vihola2012}, which automatically
tunes the covariance of a proposal density. We choose the tuning parameters
of the sampler as in \citet{Vihola2012}. For all experiments, we
simulate a total of 70,000 draws; the initial 20,000 draws are used
for warmup and the subsequent 50,000 for posterior estimates. The
initial value of $\gamma$ is randomly generated from a uniform distribution
with interval $\left(-2.5,3.5\right)$. $\boldsymbol{W}$ is initialized
to an identity matrix.

We evaluate the results of inference of $\gamma$ according to four
measures. The first is the failure rate (Fail): when the estimated
inter-quantile range of a target posterior density is larger than
1 or smaller than 0.01, we consider that the MCMC run has failed.
The second is the mean squared error of the posterior mean estimate
(MSE). The third is the inter-quantile range of the posterior density
(IQR). The fourth is the total computation time measured in seconds
(Speed). We conduct 500 experiments.

We compare the results for the precision matrix estimators. The left
halves of Tables 1-3 show the results for the standard precision matrix
estimator and the right halves show those for the NER estimator. The
upper parts of Tables 1-3 report the results for $K=50$, the middle
parts for $K=150$, and the lower parts for $K=250$. We see a similar
pattern from the tables, regardless of the number of latent factors
$S$ relative to $K$. 

When $K>N$, the number of Fails for the standard estimator exceed
half the number of experiments (500), and the posterior simulations
using the standard estimator are unsuccessful. For instance, when
using $S=3$, $K=250$, and\textit{ Random}, the standard estimator
failed 485 of 500 experiments (the last row of Table 3). In contrast,
even with $K>N$, unless using \textit{Concurrent}, the number of
Fails for the NER estimator is zero, which means that the NER estimator
provides reasonable posterior estimates. Therefore, when $K>N$, only
the NER estimator is a viable option.

For most cases, the MSEs for the NER estimator are smaller than those
for the standard estimator. For instance, when using $S=3$, $K=150$,
and\textit{ Random}, the MSE for the standard estimator was 0.0809,
while that for the NER estimator was 0.0155. Thus, in terms of estimation
accuracy, the NER estimator outperforms the standard estimator. While
the NER estimator does not have a significant advantage over the standard
estimator in terms of MSE for relative easy cases, that is, $K$ and/or
$S$ are small, even when the number of moment conditions $K$ is
smaller than the sample size $N$, the NER estimator is likely to
obtain a more accurate posterior mean estimate than the standard precision
estimator. Notably, when $K>N$, the posterior simulation using the
NER estimator is almost as precise as the cases with $K<N$. For instance,
when using $S=3$ and\textit{ Random}, the standard estimator had
MSEs of 0.247, 0.0809, and N/A (all the experiments failed) for $K=50,$
150, and 250, respectively. In contrast, for the same cases, the NER
estimator obtained MSEs of 0.0215, 0.0155, and 0.0166 for $K=50$,
150, and 250, respectively. A comparison between the results for the
\textit{Oracle} cases with different precision estimators and $K=50,150$
reveals that the NER estimator is not better than the standard one
if the true value of $\boldsymbol{\theta}$ is known. For instance,
when using $S=3$ and\textit{ Random}, MSEs for the standard estimator
are 0.0104 and 0.0012 for $K=50$ and 150, respectively, while those
for the NER estimator are 0.0185 and 0.0121 for $K=50$ and 150, respectively.
However, as suggested by a comparison between MSEs for cases using
updating procedures other than Oracle, in practical situations, the
gain from the numerical stability of the NER estimator outweighs its
efficiency loss.

\begin{table}[p]
\caption{Comparison of different approaches (1): $S=K$}

\medskip{}

\begin{centering}
\begin{tabular}{clrr@{\extracolsep{0pt}.}lr@{\extracolsep{0pt}.}lrcrccr@{\extracolsep{0pt}.}l}
\hline 
 & Estimator & \multicolumn{6}{c}{Standard} &  & \multicolumn{5}{c}{NER}\tabularnewline
\hline 
$K$ & Adaptation & Fail\ \ \  & \multicolumn{2}{c}{MSE} & \multicolumn{2}{c}{IQR} & Time &  & Fail\ \ \  & MSE & IQR & \multicolumn{2}{c}{Time}\tabularnewline
\hline 
\multirow{5}{*}{50} & Oracle & 0/500 & 0&0089 & 0&1292 & 2.2 &  & 0/500 & 0.0166 & 0.1349 & 2&2\tabularnewline
 & Concurrent & 201/500 & \multicolumn{2}{c}{\textendash{}} & \multicolumn{2}{c}{\textendash{}} & 9.9 &  & 486/500 & \textendash{} & \textendash{} & 35&9\tabularnewline
 & Stochastic & 0/500 & 0&0205 & 0&1305 & 4.6 &  & 0/500 & 0.0184 & 0.1529 & 11&2\tabularnewline
 & Continuous & 0/500 & 0&0210 & 0&1289 & 4.9 &  & 0/500 & 0.0215 & 0.1348 & 12&4\tabularnewline
 & Random & 0/500 & 0&0210 & 0&1287 & 2.3 &  & 0/500 & 0.0215 & 0.1348 & 2&6\tabularnewline
\hline 
\multirow{5}{*}{150} & Oracle & 0/500 & 0&0012 & 0&0480 & 5.7 &  & /500 & 0.0114 & 0.0939 & 5&7\tabularnewline
 & Concurrent & 383/500 & \multicolumn{2}{c}{\textendash{}} & \multicolumn{2}{c}{\textendash{}} & 56.4 &  & 500/500 & \textendash{} & \textendash{} & 247&5\tabularnewline
 & Stochastic & 0/500 & 0&0336 & 0&0689 & 19.9 &  & /500 & 0.0137 & 0.1209 & 67&8\tabularnewline
 & Continuous & 0/500 & 0&0686 & 0&0480 & 21.8 &  & /500 & 0.0162 & 0.0939 & 76&1\tabularnewline
 & Random & 0/500 & 0&0711 & 0&0481 & 6.3 &  & /500 & 0.0166 & 0.0936 & 8&3\tabularnewline
\hline 
\multirow{5}{*}{250} & Oracle & 375/500 & \multicolumn{2}{c}{\textendash{}} & \multicolumn{2}{c}{\textendash{}} & 5.4 &  & 0/500 & 0.0115 & 0.0772 & 5&4\tabularnewline
 & Concurrent & 500/500 & \multicolumn{2}{c}{\textendash{}} & \multicolumn{2}{c}{\textendash{}} & 458.5 &  & 500/500 & \textendash{} & \textendash{} & 259&1\tabularnewline
 & Stochastic & 395/500 & \multicolumn{2}{c}{\textendash{}} & \multicolumn{2}{c}{\textendash{}} & 126.1 &  & 0/500 & 0.0142 & 0.1075 & 76&4\tabularnewline
 & Continuous & 480/500 & \multicolumn{2}{c}{\textendash{}} & \multicolumn{2}{c}{\textendash{}} & 138.4 &  & 0/500 & 0.0162 & 0.0782 & 81&7\tabularnewline
 & Random & 450/500 & \multicolumn{2}{c}{\textendash{}} & \multicolumn{2}{c}{\textendash{}} & 10.9 &  & 0/500 & 0.0176 & 0.0769 & 8&6\tabularnewline
\hline 
\end{tabular}\medskip{}
\par\end{centering}
\centering{}%
\begin{minipage}[t]{0.8\columnwidth}%
Notes: The column labeled Fail reports the number of failed runs.
Column MSE reports the mean squared errors of posterior mean estimates.
Column IQR reports the inter-quantile ranges of posterior densities.
Column Time reports the averages of computation time measured in seconds.%
\end{minipage}
\end{table}

\begin{table}[p]
\caption{Comparison of different approaches (2): $S=K/2$}

\medskip{}

\begin{centering}
\begin{tabular}{clrr@{\extracolsep{0pt}.}lr@{\extracolsep{0pt}.}lr@{\extracolsep{0pt}.}lcrccr@{\extracolsep{0pt}.}l}
\hline 
 & Estimator & \multicolumn{7}{c}{Standard} &  & \multicolumn{5}{c}{NER}\tabularnewline
\hline 
$K$ & Adaptation & Fail\ \ \  & \multicolumn{2}{c}{MSE} & \multicolumn{2}{c}{IQR} & \multicolumn{2}{c}{Time} &  & Fail\ \ \  & MSE & IQR & \multicolumn{2}{c}{Time}\tabularnewline
\hline 
\multirow{5}{*}{50} & Oracle & 0/500 & 0&0085 & 0&1290 & 2&2 &  & 0/500 & 0.0151 & 0.1346 & 2&2\tabularnewline
 & Concurrent & 212/500 & \multicolumn{2}{c}{\textendash{}} & \multicolumn{2}{c}{\textendash{}} & 9&9 &  & 481/500 & \textendash{} & \textendash{} & 35&7\tabularnewline
 & Stochastic & 0/500 & 0&0217 & 0&1308 & 4&6 &  & 0/500 & 0.0179 & 0.1513 & 11&1\tabularnewline
 & Continuous & 0/500 & 0&0228 & 0&1288 & 4&9 &  & 0/500 & 0.0205 & 0.1335 & 12&3\tabularnewline
 & Random & 0/500 & 0&0228 & 0&1287 & 2&3 &  & 0/500 & 0.0204 & 0.1343 & 2&6\tabularnewline
\hline 
\multirow{5}{*}{150} & Oracle & 0/500 & 0&0012 & 0&0476 & 5&7 &  & 0/500 & 0.0112 & 0.0937 & 5&7\tabularnewline
 & Concurrent & 402/500 & \multicolumn{2}{c}{\textendash{}} & \multicolumn{2}{c}{\textendash{}} & 56&4 &  & 500/500 & \textendash{} & \textendash{} & 247&8\tabularnewline
 & Stochastic & 0/500 & 0&0380 & 0&0701 & 19&9 &  & 0/500 & 0.0135 & 0.1187 & 68&0\tabularnewline
 & Continuous & 0/500 & 0&0619 & 0&0477 & 21&7 &  & 0/500 & 0.0169 & 0.0931 & 76&2\tabularnewline
 & Random & 0/500 & 0&0673 & 0&0482 & 6&3 &  & 0/500 & 0.0160 & 0.0935 & 8&3\tabularnewline
\hline 
\multirow{5}{*}{250} & Oracle & 359/500 & \multicolumn{2}{c}{\textendash{}} & \multicolumn{2}{c}{\textendash{}} & 10&4 &  & 0/500 & 0.0095 & 0.0787 & 10&4\tabularnewline
 & Concurrent & 500/500 & \multicolumn{2}{c}{\textendash{}} & \multicolumn{2}{c}{\textendash{}} & 1036&2 &  & 500/500 & \textendash{} & \textendash{} & 757&8\tabularnewline
 & Stochastic & 398/500 & \multicolumn{2}{c}{\textendash{}} & \multicolumn{2}{c}{\textendash{}} & 226&2 &  & 0/500 & 0.0113 & 0.1076 & 200&1\tabularnewline
 & Continuous & 475/500 & \multicolumn{2}{c}{\textendash{}} & \multicolumn{2}{c}{\textendash{}} & 307&7 &  & 0/500 & 0.0130 & 0.0787 & 225&5\tabularnewline
 & Random & 467/500 & \multicolumn{2}{c}{\textendash{}} & \multicolumn{2}{c}{\textendash{}} & 21&4 &  & 0/500 & 0.0135 & 0.0782 & 18&3\tabularnewline
\hline 
\end{tabular}\medskip{}
\par\end{centering}
\centering{}%
\begin{minipage}[t]{0.8\columnwidth}%
Notes: The column labeled Fail reports the number of failed runs.
Column MSE reports the mean squared errors of posterior mean estimates.
Column IQR reports the inter-quantile ranges of posterior densities.
Column Time reports the averages of computation time measured in seconds.%
\end{minipage}
\end{table}
\begin{table}[p]
\caption{Comparison of different approaches (3): $S=3$}

\medskip{}

\begin{centering}
\begin{tabular}{clrr@{\extracolsep{0pt}.}lr@{\extracolsep{0pt}.}lr@{\extracolsep{0pt}.}lcrccr@{\extracolsep{0pt}.}l}
\hline 
 & Estimator & \multicolumn{7}{c}{Standard} &  & \multicolumn{5}{c}{NER}\tabularnewline
\hline 
$K$ & Adaptation & Fail\ \ \  & \multicolumn{2}{c}{MSE} & \multicolumn{2}{c}{IQR} & \multicolumn{2}{c}{Time} &  & Fail\ \ \  & MSE & IQR & \multicolumn{2}{c}{Time}\tabularnewline
\hline 
\multirow{5}{*}{50} & Oracle & 0/500 & 0&0104 & 0&1288 & 2&6 &  & 0/500 & 0.0185 & 0.1258 & 2&6\tabularnewline
 & Concurrent & 208/500 & \multicolumn{2}{c}{\textendash{}} & \multicolumn{2}{c}{\textendash{}} & 11&2 &  & 480/500 & \textendash{} & \textendash{} & 40&8\tabularnewline
 & Stochastic & 0/500 & 0&0242 & 0&1303 & 5&4 &  & 0/500 & 0.0207 & 0.1352 & 12&8\tabularnewline
 & Continuous & 0/500 & 0&0247 & 0&1283 & 5&7 &  & 0/500 & 0.0219 & 0.1251 & 14&2\tabularnewline
 & Random & 0/500 & 0&0247 & 0&1282 & 2&8 &  & 0/500 & 0.0215 & 0.1267 & 3&1\tabularnewline
\hline 
\multirow{5}{*}{150} & Oracle & 0/500 & 0&0012 & 0&0479 & 6&2 &  & 0/500 & 0.0121 & 0.0868 & 6&2\tabularnewline
 & Concurrent & 398/500 & \multicolumn{2}{c}{\textendash{}} & \multicolumn{2}{c}{\textendash{}} & 71&3 &  & 500/500 & \textendash{} & \textendash{} & 285&1\tabularnewline
 & Stochastic & 0/500 & 0&0379 & 0&0689 & 24&1 &  & 0/500 & 0.0139 & 0.1055 & 77&7\tabularnewline
 & Continuous & 0/500 & 0&0761 & 0&0487 & 26&5 &  & 0/500 & 0.0165 & 0.0863 & 87&3\tabularnewline
 & Random & 0/500 & 0&0809 & 0&0488 & 7&0 &  & 0/500 & 0.0155 & 0.0867 & 9&2\tabularnewline
\hline 
\multirow{5}{*}{250} & Oracle & 363/500 & \multicolumn{2}{c}{\textendash{}} & \multicolumn{2}{c}{\textendash{}} & 10&8 &  & 0/500 & 0.0110 & 0.0706 & 10&8\tabularnewline
 & Concurrent & 500/500 & \multicolumn{2}{c}{\textendash{}} & \multicolumn{2}{c}{\textendash{}} & 1157&6 &  & 500/500 & \textendash{} & \textendash{} & 829&3\tabularnewline
 & Stochastic & 480/500 & \multicolumn{2}{c}{\textendash{}} & \multicolumn{2}{c}{\textendash{}} & 295&7 &  & 0/500 & 0.0142 & 0.0937 & 218&7\tabularnewline
 & Continuous & 493/500 & \multicolumn{2}{c}{\textendash{}} & \multicolumn{2}{c}{\textendash{}} & 343&0 &  & 0/500 & 0.0161 & 0.0706 & 246&3\tabularnewline
 & Random & 485/500 & \multicolumn{2}{c}{\textendash{}} & \multicolumn{2}{c}{\textendash{}} & 23&1 &  & 0/500 & 0.0166 & 0.0707 & 19&4\tabularnewline
\hline 
\end{tabular}\medskip{}
\par\end{centering}
\centering{}%
\begin{minipage}[t]{0.8\columnwidth}%
Notes: The column labeled Fail reports the number of failed runs.
Column MSE reports the mean squared errors of posterior mean estimates.
Column IQR reports the inter-quantile ranges of posterior densities.
Column Time reports the averages of computation time measured in seconds.%
\end{minipage}
\end{table}

We also investigate the sensitivity of the above results to the choice
of split location $N^{*}$. We conduct Monte Carlo experiments using
different $N^{*}$ and two preferred adaptation strategies, \textit{Stochastic}
and \textit{Random}. Following \citet{Lam2016}, we consider the grid
of (\ref{eq: grid bigN_star}) (each $N^{*}$ is rounded to the nearest
integer). Table 4 shows that the NER estimator consistently outperforms
the standard estimator, irrespective of the split location choice.
In terms of MSE, a moderate value of $N^{*}$ is preferred. To investigate
if this result is in agreement with the criteria based on the Frobenius
norm (\ref{eq: fro-norm crit}), we simulate the values of (\ref{eq: fro-norm crit})
for different random permutations of the moment conditions using the
true parameter. Panel (a) of Figure 1 reports the median and 90 percentile
intervals of the simulated values for a fine grid $\left\{ 0.1N,\;0.15N,...,0.9N\right\} $.
We only report the results for $K=250$, as those for $K=50,150$
are qualitatively similar. As evident from the panel, an extremely
high $N^{*}$ is not preferred, but the criterion is not sufficiently
informative to select a good $N^{*}$ from a considerably large range.
The variability of the criterion is not attributable to the small
sample size. We conduct the same simulation as in panel (a) but the
sample size increases to $N=5,000$. Panel (b) of Figure 1 shows the
results. As is the case of $N=200$, the values of the criterion based
on the Frobenius norm are almost indifferent for a large range. Therefore,
we recommend setting $N^{*}$ to approximately half the sample size
as default.

\begin{table}[t]
\caption{Comparison of different choices of $N^{*}$}

\medskip{}

\begin{centering}
\begin{tabular}{lllccccc}
\hline 
 &  & Adaptation & \multicolumn{2}{c}{Stochastic} &  & \multicolumn{2}{c}{Random}\tabularnewline
\hline 
$K$ & Estimator & $N^{*}$ & MSE & IQR &  & MSE & IQR\tabularnewline
\hline 
\multirow{8}{*}{50} & Standard &  & 0.0242 & 0.1303 &  & 0.0247 & 0.1282\tabularnewline
\cline{2-8} 
 & \multirow{7}{*}{NER} & $28\left(=\left[2N^{1/2}\right]\right)$ & 0.0223 & 0.1174 &  & 0.0237 & 0.1109\tabularnewline
 &  & $40\left(=0.2N\right)$ & 0.0220 & 0.1209 &  & 0.0225 & 0.1131\tabularnewline
 &  & $80\left(=0.4N\right)$ & 0.0211 & 0.1299 &  & 0.0229 & 0.1201\tabularnewline
 &  & $120\left(=0.6N\right)$ & 0.0207 & 0.1354 &  & 0.0220 & 0.1261\tabularnewline
 &  & $160\left(=0.8N\right)$ & 0.0206 & 0.1382 &  & 0.0218 & 0.1266\tabularnewline
 &  & $164\left(=\left[N-2.5N^{1/2}\right]\right)$ & 0.0205 & 0.1386 &  & 0.0223 & 0.1252\tabularnewline
 &  & $178\left(=\left[N-1.5N^{1/2}\right]\right)$ & 0.0206 & 0.1408 &  & 0.0239 & 0.1233\tabularnewline
\hline 
\multirow{8}{*}{150} & Standard &  & 0.0379 & 0.0689 &  & 0.0809 & 0.0488\tabularnewline
\cline{2-8} 
 & \multirow{7}{*}{NER} & $28\left(=\left[2N^{1/2}\right]\right)$ & 0.0171 & 0.0854 &  & 0.0182 & 0.0733\tabularnewline
 &  & $40\left(=0.2N\right)$ & 0.0163 & 0.0904 &  & 0.0177 & 0.0753\tabularnewline
 &  & $80\left(=0.4N\right)$ & 0.0145 & 0.1016 &  & 0.0165 & 0.0816\tabularnewline
 &  & $120\left(=0.6N\right)$ & 0.0139 & 0.1055 &  & 0.0160 & 0.0864\tabularnewline
 &  & $160\left(=0.8N\right)$ & 0.0143 & 0.1051 &  & 0.0161 & 0.0887\tabularnewline
 &  & $164\left(=\left[N-2.5N^{1/2}\right]\right)$ & 0.0144 & 0.1054 &  & 0.0158 & 0.0882\tabularnewline
 &  & $178\left(=\left[N-1.5N^{1/2}\right]\right)$ & 0.0147 & 0.1083 &  & 0.0165 & 0.0847\tabularnewline
\hline 
\multirow{8}{*}{250} & Standard &  & \textendash{} & \textendash{} &  & \textendash{} & \textendash{}\tabularnewline
\cline{2-8} 
 & \multirow{7}{*}{NER} & $28\left(=\left[2N^{1/2}\right]\right)$ & 0.0177 & 0.0743 &  & 0.0184 & 0.0591\tabularnewline
 &  & $40\left(=0.2N\right)$ & 0.0168 & 0.0796 &  & 0.0187 & 0.0611\tabularnewline
 &  & $80\left(=0.4N\right)$ & 0.0148 & 0.0913 &  & 0.0173 & 0.0667\tabularnewline
 &  & $120\left(=0.6N\right)$ & 0.0142 & 0.0938 &  & 0.0163 & 0.0707\tabularnewline
 &  & $160\left(=0.8N\right)$ & 0.0148 & 0.0916 &  & 0.0166 & 0.0710\tabularnewline
 &  & $164\left(=\left[N-2.5N^{1/2}\right]\right)$ & 0.0149 & 0.0916 &  & 0.0175 & 0.0724\tabularnewline
 &  & $178\left(=\left[N-1.5N^{1/2}\right]\right)$ & 0.0153 & 0.0939 &  & 0.0176 & 0.0693\tabularnewline
\hline 
\end{tabular} 
\par\end{centering}
\medskip{}

\centering{}%
\begin{minipage}[t]{0.8\columnwidth}%
Notes: The column labeled Fail reports the number of failed runs.
Column MSE reports the mean squared errors of posterior mean estimates.
Column IQR reports the inter-quantile ranges of posterior densities. %
\end{minipage}
\end{table}

\begin{figure}[p]
\caption{The Frobenius norm criterion for different permutations}

\medskip{}
\begin{minipage}[t]{0.45\columnwidth}%
\begin{center}
(a) $N=200$ 
\par\end{center}
\includegraphics[clip,scale=0.45]{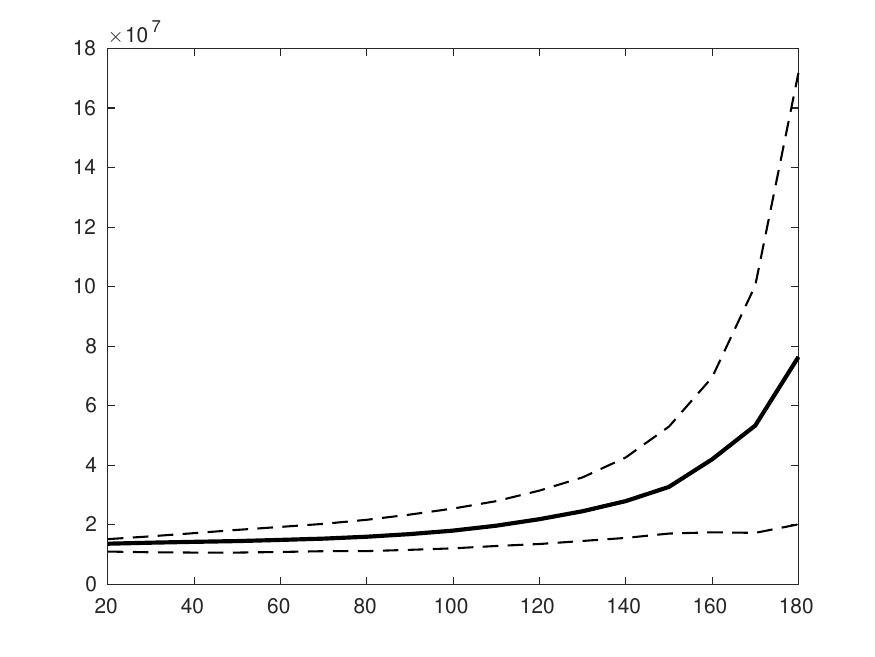}%
\end{minipage}\hfill{}%
\begin{minipage}[t]{0.45\columnwidth}%
\begin{center}
(b) $N=5000$
\par\end{center}
\includegraphics[clip,scale=0.45]{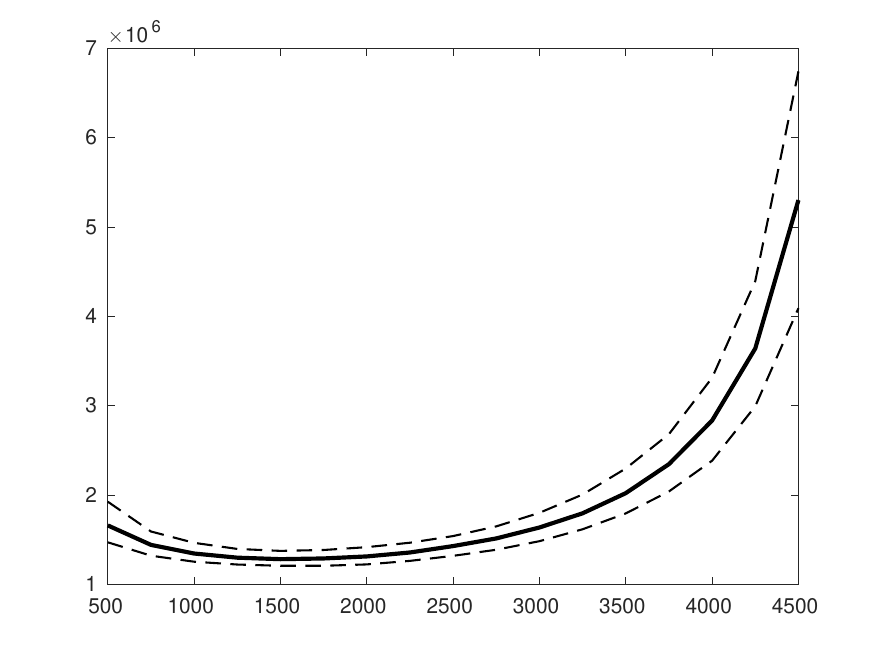}%
\end{minipage}\medskip{}

\centering{}%
\begin{minipage}[t]{0.8\columnwidth}%
Notes: Solid lines denote the median, and dashed lines denote the
90 percentile interval. $K=250$. Moment conditions are calculated
based on the true parameter value.%
\end{minipage}
\end{figure}

Next, we compare the results of the adaptation strategies. There are
five points worth mentioning. First, \textit{Concurrent} does not
work despite high computational cost. Second, the relative advantage
of \textit{Stochastic} to \textit{Concurrent} in terms of numerical
stability is in line with \citet{Yin2011}. Third, in terms of MSE,
all \textit{Stochastic}, \textit{Continuous,} and \textit{Random}
work well. \textit{Stochastic} is better than \textit{Continuous}
and \textit{Random}, while \textit{Continuous} and \textit{Random}
are comparable. Fourth, as the IQR estimates show, \textit{Continuous}
and \textit{Random} are more optimistic than \textit{Stochastic}.
Fifth, \textit{Random} is much faster than \textit{Stochastic} and
\textit{Continuous}. Figure 2 provides a typical example of recursive
posterior mean and the occurrence of random adaptation (NER estimator,
$K=150$). From this figure, a posterior mean is fairly fast to converge,
which indicates that most updates of the weighting matrix in \textit{Continuous}
are essentially redundant. We find \textit{Random} has a good balance
between statistical and computational efficiency; therefore, it is
recommendable for a test run. Although \textit{Stochastic} is computationally
demanding, it is more accurate and conservative than \textit{Random}.
Therefore, it is suitable for a final estimate. 

\begin{figure}[p]
\caption{An example of random adaptation}
\medskip{}

\includegraphics{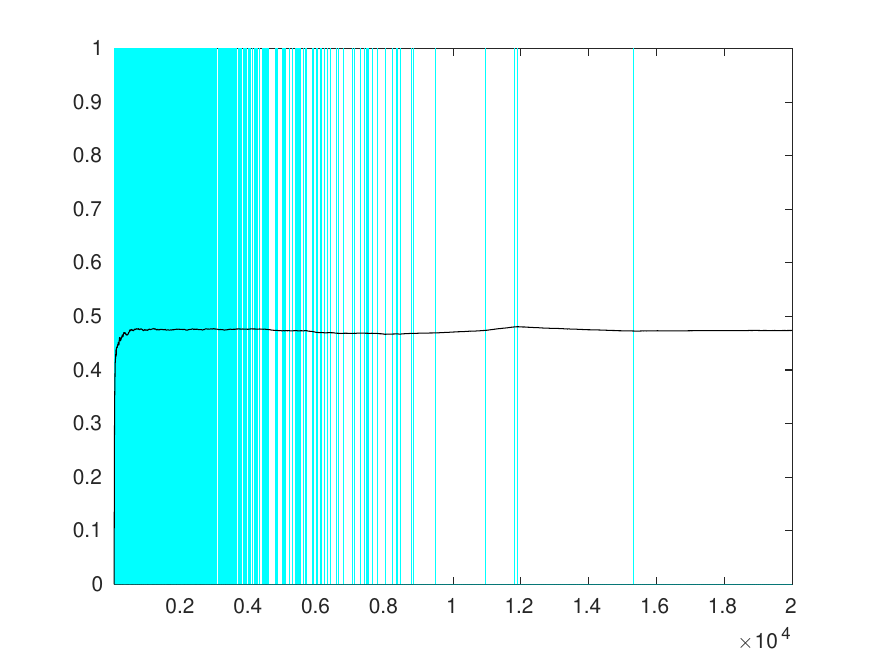}

\medskip{}

\centering{}%
\begin{minipage}[t]{0.8\columnwidth}%
Notes: The x-axis denotes MCMC iterations and the y-axis denotes parameter
values. A thin solid vertical line denotes the occurrence of adaptation.
A bold solid line denotes a recursive mean of posterior samples.%
\end{minipage}
\end{figure}

\section{Application}

To demonstrate the proposed method, we apply it to a demand analysis
for automobiles. \citet{Berry1995} consider an IV regression model
of demand for automobiles specified by
\[
y_{i,t}=\gamma p_{i,t}+\boldsymbol{\delta}^{\top}\boldsymbol{x}_{i,t}+u_{i,t},
\]
\[
y_{i,t}=\log\left(s_{i,t}\right)-\log\left(s_{0,t}\right).
\]
$s_{i,t}$ denotes the market share of product $i$ on market $t$,
with subscript $0$ denoting the outside option. The treatment $p_{i,t}$
is the product price. $u_{i,t}$ is an error term, and $\gamma$ and
$\boldsymbol{\delta}$ are the parameters to be estimated. The primary
focus of this application is the inference of $\gamma$.

We consider two specifications.\footnote{All data are extracted from R package \texttt{hdm} (version 0.2.3). }
The first specification coincides with \citet{Berry1995} as follows:
A vector of covariates $\boldsymbol{x}_{n}$ includes four covariates,
namely, air conditioning dummy, horsepower to weight ratio, miles
per dollar, and vehicle size. A set of instruments contains the four
covariates and ten variables, namely, the sum of each covariate taken
across models made by product $t$'s firm, the sum of each covariate
taken across competitor firms' products, the total number of models
produced by product $t$'s firm, and the total number of models produced
by the firm's competitors. The second specification is an extension
of the first, which is considered in \citet{Chernozhukov2015}. $\boldsymbol{x}_{n}$
and $\boldsymbol{z}_{n}$ extend from the first case by incorporating
a time trend, quadratic and cubic terms of all continuous covariates,
and first-order interaction terms. The numbers of the instruments
in the first and second specifications are 10 and 48, respectively.
The sample size is $N=2,217$, being larger than the numbers of instruments.
Nevertheless, because of the ill-posedness of the data set, the covariance
of a classical estimator is nearly singular. We use a constant prior;
thus, if the relationship between the instruments and the treatment
is linear and the distributions of residuals are normal, a posterior
estimate coincides with a two-stage least square estimate. The posterior
estimate is obtained using different combinations of precision matrix
estimators and the adaptation of proposal density. We sample a total
of 250,000 posterior draws with the last 200,000 drawn for posterior
analysis. 

Table 5 summarizes the results of the posterior estimate for the coefficient
on price. Although the number of moment conditions is fairly smaller
than the sample size, MCMC runs using \textit{Concurrent }fails to
converge. By contrast, MCMC runs using the NER estimator obtain sensible
posterior samples, irrespective of the adaptation strategy. For comparison,
Table 5 also includes the estimates obtained using four alternative
methods. The first two are conventional: ordinary least squares (OLS)
and two-stage least squares (2SLS) methods. The second two are state-of-the-art:
IV with instrument selection based on a least absolute shrinkage and
selection operator \citep{Chernozhukov2015} (LASSO-IV), and Bayesian
IV with a factor shrinkage prior \citep{Hahn2018} (HS-IV). LASSO-IV
is designed to select fewer relevant instruments, while HS-IV is designed
to compress observed information into few latent factors. The two
methods assume a linear relationship between instruments and the endogenous
variable and Gaussianity of the error terms, while our method does
not impose such assumptions. These alternative methods obtain larger
estimates than the conventional ones, and the estimates depend significantly
on a set of (potential) instruments. By contrast, our method estimates
the coefficient to be intermediate between OLS and 2SLS, nearly irrespective
of the choice of instruments. As Figure 3 shows, the posterior densities
of $\gamma$ for alternative approaches (excluding \textit{Concurrent}
adaptation) are quite similar.

\begin{table}[t]
\caption{Posterior estimates of $\gamma$}

\medskip{}

\begin{centering}
\begin{tabular}{llr@{\extracolsep{0pt}.}lr@{\extracolsep{0pt}.}lr@{\extracolsep{0pt}.}lr@{\extracolsep{0pt}.}lr@{\extracolsep{0pt}.}lr@{\extracolsep{0pt}.}lr@{\extracolsep{0pt}.}l}
 &  & \multicolumn{6}{c}{Standard} & \multicolumn{2}{c}{} & \multicolumn{6}{c}{NER}\tabularnewline
$K$ &  & \multicolumn{2}{c}{Mean} & \multicolumn{2}{c}{Std} & \multicolumn{2}{c}{Time} & \multicolumn{2}{c}{} & \multicolumn{2}{c}{Mean} & \multicolumn{2}{c}{Std} & \multicolumn{2}{c}{Time}\tabularnewline
\hline 
\multirow{7}{*}{10} & Concurrent & \multicolumn{2}{c}{\textendash{}} & \multicolumn{2}{c}{\textendash{}} & 1214&4 & \multicolumn{2}{c}{} & \multicolumn{2}{c}{\textendash{}} & \multicolumn{2}{c}{\textendash{}} & 1213&6\tabularnewline
 & Stochastic & -0&120 & 0&049 & 334&3 & \multicolumn{2}{c}{} & -0&117 & 0&051 & 508&3\tabularnewline
 & Continuous & -0&122 & 0&051 & 363&3 & \multicolumn{2}{c}{} & -0&106 & 0&051 & 439&9\tabularnewline
 & Random & -0&122 & 0&051 & 215&7 & \multicolumn{2}{c}{} & -0&110 & 0&050 & 220&8\tabularnewline
\cline{2-16} 
 & OLS & -0&089 & 0&004 & \multicolumn{2}{c}{} & \multicolumn{2}{c}{} & \multicolumn{2}{c}{} & \multicolumn{2}{c}{} & \multicolumn{2}{c}{}\tabularnewline
 & 2SLS & -0&142 & 0&012 & \multicolumn{2}{c}{} & \multicolumn{2}{c}{} & \multicolumn{2}{c}{} & \multicolumn{2}{c}{} & \multicolumn{2}{c}{}\tabularnewline
 & LASSO-IV & -0&185 & 0&014 & \multicolumn{2}{c}{} & \multicolumn{2}{c}{} & \multicolumn{2}{c}{} & \multicolumn{2}{c}{} & \multicolumn{2}{c}{}\tabularnewline
\hline 
\multirow{6}{*}{48} & Concurrent & \multicolumn{2}{c}{\textendash{}} & \multicolumn{2}{c}{\textendash{}} & 2606&9 & \multicolumn{2}{c}{} & \multicolumn{2}{c}{\textendash{}} & \multicolumn{2}{c}{\textendash{}} & 3711&2\tabularnewline
 & Stochastic & -0&116 & 0&011 & 1230&7 & \multicolumn{2}{c}{} & -0&119 & 0&014 & 1613&8\tabularnewline
 & Continuous & -0&117 & 0&010 & 1071&4 & \multicolumn{2}{c}{} & -0&117 & 0&010 & 1432&4\tabularnewline
 & Random & -0&117 & 0&010 & 698&3 & \multicolumn{2}{c}{} & -0&119 & 0&010 & 705&6\tabularnewline
\cline{2-16} 
 & LASSO-IV & -0&221 & 0&015 & \multicolumn{2}{c}{} & \multicolumn{2}{c}{} & \multicolumn{2}{c}{} & \multicolumn{2}{c}{} & \multicolumn{2}{c}{}\tabularnewline
 & HS-IV & -0&275 & 0&018 & \multicolumn{2}{c}{} & \multicolumn{2}{c}{} & \multicolumn{2}{c}{} & \multicolumn{2}{c}{} & \multicolumn{2}{c}{}\tabularnewline
\hline 
\end{tabular} 
\par\end{centering}
\medskip{}

\centering{}%
\begin{minipage}[t]{0.8\columnwidth}%
Notes: The column labeled Mean reports mean estimates. Column Std
reports standard errors. Column Time reports computation time measured
in seconds.%
\end{minipage}
\end{table}

\begin{figure}[p]
\caption{Posterior distribution of $\gamma$}

\medskip{}

\includegraphics[clip]{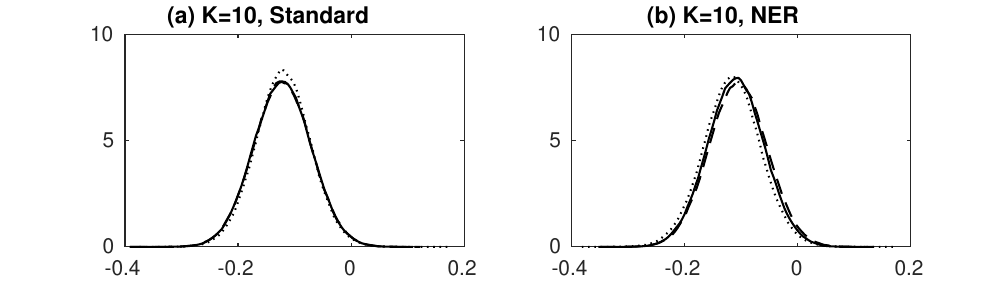} 

\medskip{}

\includegraphics[clip]{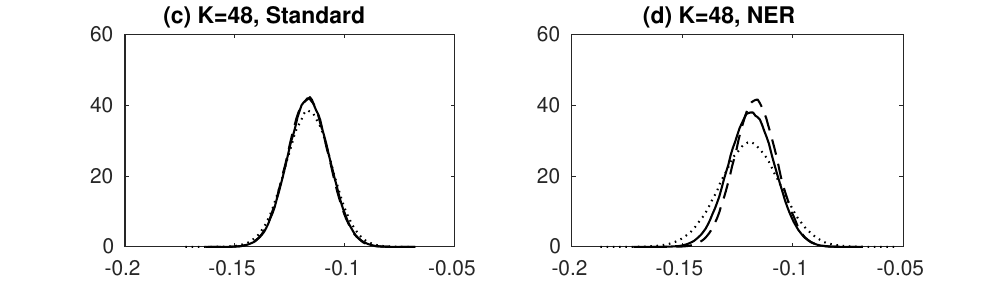}

\medskip{}

\centering{}%
\begin{minipage}[t]{0.8\columnwidth}%
Notes: Solid lines trace the mean estimates for \textit{Stochastic}.
Dashed lines trace the mean estimates for \textit{Continuous}. Dotted
lines trace the mean estimates for \textit{Random}.%
\end{minipage}
\end{figure}

\section{Discussion}

We propose a new adaptive MCMC approach to infer Bayesian GMM with
many moment conditions. Our proposal consists of two elements. The
first is the use of a nonparametric eigenvalue-regularized precision
matrix estimator \citep{Lam2016} for estimating the weighting matrix.
This prevents us from ill-estimating the weighting matrix. The second
is the use of random adaptation. Setting adaptation probability as
exponentially decreasing can significantly reduce the computational
burden, while retaining statistical efficiency. We demonstrate the
superiority of the proposed approach over existing approaches through
simulation, and by applying it to a demand analysis for automobiles.

Several promising research areas stem from this study. First, a theoretical
investigation of the effects of tuning/estimation of a weighting matrix
on the posterior density is necessary, which is absent in the literature.
Second, while the proposed approach seems to be fairly robust to $N^{*}$,
there is room for improvement by finding a better $N^{*}$. Third,
while this study only addresses problems caused by many moment conditions,
it is also important to solve the problems caused by many unknown
parameters. The proposed method should serve as a stepping stone for
 the further development of inferential methods for high-dimensional
Bayesian GMM. Finally, it is worth conducting a thorough comparison
between the proposed approach and existing classical and Bayesian
approaches tailored to a specific class of models, such as IV regressions
and dynamic panel models. 

\bibliographystyle{jasalike}
\bibliography{Reference}

\end{document}